\documentclass[10pt,aps,showpacs,twocolumn,unsortedaddress,nofootinbib,floatfix]{revtex4-1}
\usepackage[dvipsnames]{xcolor}
\usepackage{graphicx}
\usepackage{bm}
\usepackage{physics}
\usepackage[version=4]{mhchem}
\usepackage{mathtools,mathrsfs,amsfonts,dsfont}
\usepackage{relsize}
\usepackage{scalerel}
\usepackage{todonotes}
\usepackage{ulem}
\usepackage{xcolor}
\usepackage[unicode=true, pdfusetitle, bookmarks=true, bookmarksnumbered=false, bookmarksopen=false, breaklinks=true, pdfborder={0 0 0}, backref=false, colorlinks=true, linkcolor=blue, citecolor=blue, urlcolor=blue]{hyperref}
\usepackage{enumitem}
\usepackage{cancel}
\allowdisplaybreaks

\newcommand{\Ex}{\sm\vu{x}}
\newcommand{\Ey}{\sm\vu{y}}
\newcommand{\Ez}{\sm\vu{z}}
\newcommand{\sm}{\kern0.1em}
\newcommand{\smalldiv}{\raisebox{-0.2ex}{\resizebox{!}{1.6ex}{\sm/\sm}}}

\DeclareMathAlphabet{\mathbbold}{U}{bbold}{m}{n}

\begin{document}

\title{Relativistic electron wave packets featuring persistent quantum backflow}

\author{Siddhant Das}
\affiliation{Mathematisches Institut, Ludwig-Maximilians-Universit\"at M\"unchen, Theresienstr.\ 39, D-80333 M\"unchen, Germany}
\date{Dec.\ 25, 2021}

\begin{abstract}
Closed-form, normalizable solutions of Dirac's equation propagating within a semi-infinite cylindrical waveguide are obtained in terms of ordinary and modified Bessel functions. These relativistic wave packets induce quantum backflow on a cross-section of the cylinder at practically any distance along the waveguide, becoming spin-polarized in the nonrelativistic limit. The predicted backflow is stable in time and is manifest regardless of the initial wave function.
\end{abstract}

\maketitle
\normalem

Dirac's wave equation 
\begin{equation}\label{Deq}
    i\hbar\left(\frac{\partial}{\partial t}+c\,\bm{\alpha}\kern-0.15em\cdot\kern-0.16em\bm{\nabla}\right)\kern-0.1em\psi=m c^2\beta\sm\psi,
\end{equation}
forms the basis of a relativistic quantum theory of electrons and positrons. \(\psi(\vb{r},t)\) is a bispinor wave function and \(\smash{\bm{\alpha}=(\alpha_x,\alpha_y,\alpha_z)}\), \(\beta\) are the Dirac matrices, taken here in their standard representation with \(\beta\) diagonal. The conserved probability and probability current/flux densities associated with \eqref{Deq}, \(\smash{\varrho(\vb{r},t)=\psi^\dagger\psi}\) and
\begin{equation}\label{Dcurrent}
     \vb{j}(\vb{r},t)=\psi^\dagger\bm{\alpha}\sm\psi
\end{equation}
(\(\psi^\dagger\) being the Hermitian adjoint spinor), provide a basis for the probabilistic interpretation of \(\psi\) \`a la Born.

Contra nonrelativistic wave mechanics, such an interpretation, if pushed too far, leads to inconsistencies even for freely moving particles, see \cite[p.\ 32]{thaller}, \cite[Sec.\ 8.5, Ch.\ 13]{GreinerWaveEquations}. Therefore, it is unclear whether one should view the Dirac equation as a single-particle wave equation like the Schr\"odinger equation. However, it is widely acknowledged to describe quantum phenomena ``where velocities are so high that relativistic kinematical effects are measurable, but where the energies are sufficiently small that pair creation occurs with negligible probability'' \cite[p.\ VI]{thaller}. Within this not so sharply defined, yet very broad domain of applicability, exact solutions of \eqref{Deq} are of great value to the physicist when applied with some discretion \cite{comment}.

While non-normalizable solutions are well-known \cite{Birula,Ziolkowski,Hillion}, normalizable (i.e., square-integrable) ones in 3+1 spacetime dimensions are few and far between \cite{hopfian,hopfianA}, \cite[p.\ 502]{Lu}. In particular, normalizable wave packets (WPs) propagating in specified electrodynamic potentials are unknown with the exception of \emph{stationary} bound states described, for example, in \cite{hydrogen,3Dbox,2cntr}, \cite[p.\ 217]{GreinerWaveEquations}, or \cite[p.\ 4]{ingenius}. On the other hand, the natural length and time scales of Eq.\ \eqref{Deq}, the Compton wavelength \(\smash{\lambda\sim 10^{-12}\sm\text{m}}\) and the reciprocal zitterbewegung-frequency \(\smash{\sim 10^{-22}\sm\text{s}}\), respectively, are so incredibly small that numerical simulations are not only computationally expensive but also beset with the so-called Fermion doubling problem \cite{Fermiondoubling}. Thus, there is a continued interest in analytic solutions.

\begin{figure}[!ht]
    \centering
    \includegraphics[width=0.85\columnwidth]{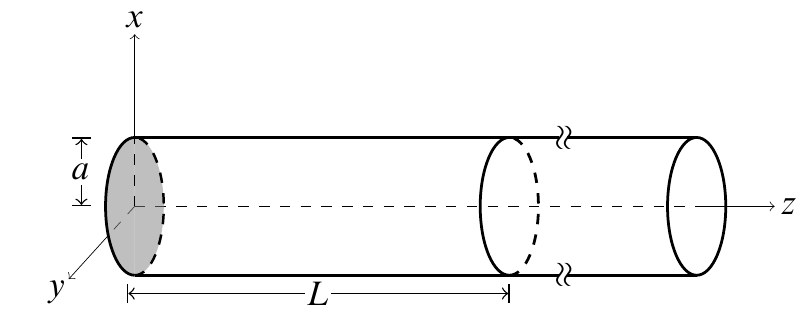}
    \caption{A semi-infinite cylindrical waveguide of radius \(a\) mounted on the \(xy\)-plane of a right-handed coordinate system, the axis of the cylinder defining the \(z\)-axis.}
    \label{WWGG}
\end{figure}

In this letter, we obtain a few normalizable \emph{positive energy} solutions of Eq.\ \eqref{Deq} that describe relativistic electrons propagating within a semi-infinite waveguide (WG), depicted in Fig.\ \ref{WWGG}. These WPs are localized initially in the vicinity of the \(xy\)-plane, the end face of the WG, permeating the entire WG over time and provoking quantum backflow (BF) on a cross-section of the cylinder at distances \(\smash{L\gg\lambda}\) and times \(\smash{>\!L\smalldiv c}\). In the nonrelativistic limit they become spin-polarized (i.e., space-spin factorized). The predicted BF is in a sense \emph{maximal} for WPs polarized perpendicular to the WG-axis.

While the ``quantum BF effect'' (loosely understood as the flow of probability current against the direction of propagation) is of considerable interest \cite{Backflow,BF0,BF,BF2,BF3,GoussevPRR,Goussev,blackhole}, \emph{steady} BF situations are extremely difficult to achieve due to certain limitations implied by scattering theory (see below). As a result, most examples display BF for, at best, fleeting moments of time, and preclude experimental inspection FAPP (see, however, \cite{MugaBEC,MugaBEC1}). Besides, the examples of quantum BF arising from Dirac's equation have so far been limited to plane waves in one spatial dimension \cite{Dirac1d,BMrelativistic}. The non-normalizability of such states notwithstanding, the restriction to a single spatial dimension compromises on both the representation of spin-structure, as well as the vectorial character of the probability current that is central to the discussion of BF.

In view of this, the \emph{stable}, \emph{spin-dependent} BF observable in the present set-up is one-of-its-kind and experimentally very promising. It entails striking consequences, e.g., in arrival-time or time-of-flight experiments \cite{DD,SciAm} amenable to present-day Paul trap technology \cite{Haffner13,Haffner17,Haffner20,Nature2021,*Nature2021applied}. (The importance of BF in arrival-time experiments has long been emphasized \cite{BohmbeatsKij,LeavOTS,*LeavensNL,Delgado,Vona}.) In this regard, the present work also provides for extending the considerations of \cite{DD,Exotic} to relativistic regimes.

\paragraph*{Waveguide model:} In what follows, we employ cylindrical polar coordinates \((\rho,\phi,z)\). The separability of Dirac's equation in axially symmetric potentials has been widely studied \cite{Greinercylinder,Leviatan}, \cite[p.\ 134]{BG}, even with a non-minimal (or anomalous moment) coupling \cite{Vilalbaneutral,DiracPauli,HoRoy}, driven by applications in, e.g., electron vortex beam experiments \cite{vortexbeam,Bvortex,Barnett,pulsevortex,*pulsevortexError}, Kumakhov (or channeling) radiation studies \cite{channelingGreiner,paraxial,RadiationReview}, the Aharonov-Bohm (-Casher) effect, and neutron confinement \cite{Bakkesame}. There are only two axisymmetric (electrostatic) potentials for which the Dirac equation is solvable, \emph{viz.}, the constant potential and the \(-1\smalldiv\rho\) potential \cite{ViSh,F11,schsolvable}. The former opens the possibility of piecewise-constant potentials capable of modelling a cylindrical WG, e.g., the finite square well \cite[Sec.\ 5]{NJP} or the attractive \(\delta\)-shell \(\smash{\propto -\,\delta(\rho-a)}\) \cite{deltashell,delta2}. These, however, give rise to intractable transcendental equations due to interface/boundary conditions enforcing regularity of the solutions, hence are of little value in constructing WPs.

This is avoided if the radial potential has only one piece: besides the \(-1\smalldiv\rho\) potential cited above, two more examples, \emph{viz.}, \(-\,\delta(\rho)\) (say, a \(\delta\)-shell in the limit \(\smash{a\to0}\)), and the hard-wall (or infinite square well) potential, come to mind. Among these, the \(\delta(\rho)\) example is ill-posed \cite{Nagomi}. Propagation in the \(-1\smalldiv\rho\) potential will be discussed in a future communication. Here, we discuss WP solutions for a hard-walled WG.

By hard-walled WG we mean an impenetrable surface that traps the wave function within the semi-infinite cylinder of Fig.\ \ref{WWGG}, such that the integral of \(\varrho(\vb{r},t)\) over the WG-volume is preserved in time. To achieve this it suffices to require the normal component of the Dirac current, Eq.\ \eqref{Dcurrent}, to vanish on the WG surface comprised of the circular cylinder \(\smash{\rho=a}\), and the \(xy\)-plane \(\smash{z=0}\). That is, we must impose the conditions
\begin{equation}\label{BC}
    \psi^\dagger\alpha_\rho\psi\big|_{\rho=a}=\psi^\dagger\alpha_z\psi\big|_{z=0}=0,
\end{equation}
on the wave function, where \(\alpha_\rho=\alpha_x\cos\phi+\alpha_y\sin\phi\). 

Equation \eqref{BC} is typically implemented by means of a linear boundary condition (BC), for which there are various candidates \cite{BCBC}. The simplest possibility is the Dirichlet BC, \(\smash{\psi=0}\), on the surface of the WG. But this must be dismissed as it requires the solution to vanish everywhere \cite[p.\ 317]{BC}, unlike in nonrelativistic quantum mechanics.

Another hard-wall BC suggested in the literature is the so-called MIT bag BC: \(\smash{(\bm{\alpha}\cdot\vu{\vb{n}})\sm\psi=i\beta\psi}\) \cite{Weisskopf,MITBagMathPhys}, where \(\vu{\vb{n}}\) is the unit outward-pointing normal vector of the surface. This BC has been used for particle trapping in one-dimensional \cite{OneDbox} and three-dimensional rectangular (spherical) boxes \cite{3Dbox} (\cite[pp.\ 713-718]{3DSbox}), and for propagation in twisted tubes \cite{twisted}, etc., but could not be consistently applied for a cylindrical box \cite[App.\ A]{ShankarGreiner}. This forced the authors of \cite{ShankarGreiner} to apply different hard-wall BCs on the walls and edges of their cylindrical box, leading them to the conclusion that the MIT bag BC ``will likewise overdetermine the problem for a cubic [box], as well as for all other [box] surfaces which need more than one equi-coordinate surface to close them'' \cite[p.\ 330]{ShankarGreiner}. Although this conclusion is incorrect, given \cite{3Dbox}, it is evident that the MIT bag BC is sensitive to the geometry of the confining surface and is therefore not quite so compelling.

The chiral bag BC, which includes the MIT bag BC as a special case, is yet another hard-wall BC invoked in applications \cite{chiral}. Note, however, that the bag BCs interlace the upper (large) and lower (small) components of \(\psi\), making the nonrelativistic limit somewhat problematic.

From a theoretical viewpoint, it is unclear whether one should prefer one BC over another, insofar as they yield the expected wave functions in the nonrelativistic limit keeping the Dirac Hamiltonian \(-\sm i\hbar\sm c\,\bm{\alpha}\cdot\bm{\nabla}+mc^2\beta\) self-adjoint. For concreteness, we employ here the hard-wall BC \cite[p.\ 317]{BC}
\begin{equation}\label{BBC}
    (\mathds{1}+\beta)\sm\psi=0,
\end{equation}
which achieves these goals by forcing \emph{only} the upper (large) components of \(\psi\) to vanish on the boundaries of the WG, as opposed to, e.g., the Dirichlet BC that yields only the trivial solution. That \eqref{BBC} implies \eqref{BC} is readily seen:
\begin{equation*}
    \psi^\dagger\alpha\psi\overset{\eqref{BBC}}{=}-\sm\big(\psi^\dagger\beta\big)\sm\alpha\psi=\psi^\dagger\alpha\beta\psi\overset{\eqref{BBC}}{=}\psi^\dagger\alpha(-\sm\psi), 
\end{equation*}
given \(\beta\) is Hermitian, and \(\smash{\alpha\beta=-\beta\alpha}\) for \emph{any} \(\alpha\)-matrix. Thus, \(\smash{\psi^\dagger\alpha\psi=0}\) under \eqref{BBC}. (See \cite[Sec.\ 3]{BC}, \cite{Menon,kicked,cosmicstring} for some applications of the above BC to stationary state problems.)  

We proceed next to solving the free Dirac equation \eqref{Deq} with BC \eqref{BBC} applied on the surfaces \(\smash{\rho=a}\) and \(\smash{z=0}\), arriving at a general class of WP solutions, Eq.\ \eqref{explicit}, below. Motivating the notion of quantum BF appropriate for a three-dimensional setting, the spin-polarization dependent BF induced by these WPs is then elucidated. Finally, a few explicit examples of such WPs are presented. 

\paragraph*{Mathematical preliminaries:} We begin with the known positive energy, odd-parity plane-wave solutions of Eq.\ \eqref{Deq} \cite[Eq.\ (2.2)]{ShankarGreiner},
\begin{subequations}\label{oddparity}
\begin{align}
    \ket{\uparrow,k}&=\mqty(J_0(k_\perp\sm\rho)\sin(k z)\sm\mqty(1\\[2pt]0)\\[10pt]\displaystyle\frac{i\hbar\sm c}{E+mc^2}\sm\mqty(-\, k\sm J_0(k_\perp\sm\rho)\cos(k z)\\[2pt] k_\perp J_1(k_\perp\sm\rho)\sin(kz)\sm e^{i\phi}))\sm e^{-\sm iEt/\hbar},\\[10pt]
   \ket{\downarrow,k}&=\mqty(J_0(k_\perp\sm\rho)\sin(k z)\sm\mqty(0\\[2pt]1)\\[10pt]\displaystyle\frac{i\hbar\sm c}{E+mc^2}\sm\mqty(k_\perp J_1(k_\perp\sm\rho)\sin(kz)\sm e^{-\sm i\phi}\\[2pt] k\sm J_0(k_\perp\sm\rho)\cos(kz)))\sm e^{-\sm iEt/\hbar},
\end{align}
\end{subequations}
as our building blocks, denoted with a slight abuse of the Dirac bra-ket notation \cite{notation}. Here, \(J_\nu\) is the Bessel function of order \(\nu\), \(k_\perp\) and \(k\) are real parameters in terms of which
\begin{equation}
     E = mc^2\sqrt{1+\lambdabar^2\big(k_\perp^2+k^2_{\phantom{\perp}}\big)},
\end{equation}
and \(\smash{\lambdabar=\hbar\smalldiv(mc)}\) is the ``reduced'' Compton wavelength. The large components of these states constitute the usual nonrelativistic spin doublet with a common radial wave function, orbital angular momentum projection \(\smash{m_\ell=0}\), and spin projection \(\smash{m_s=\pm\sm1\smalldiv2}\). 

We chose the odd-parity solutions \cite{parity} because their large components \(\smash{\propto\sin(kz)}\) automatically vanish at \(\smash{z=0}\), fulfilling \eqref{BBC} (in turn, \eqref{BC}) on the end face of the WG. In order to satisfy the radial BC, the large components are made to vanish at \(\smash{\rho=a}\) by requiring
\begin{equation}\label{rbc}
    J_0(k_\perp\sm a)=0.
\end{equation}
It follows that \(\smash{k_\perp = j_{0,\sm n}\smalldiv a}\), where \(j_{0,\sm n}\) denotes the \(n\)\textsuperscript{th} zero of \(J_0\). We will consider only the \(\smash{n=1}\) mode for brevity, in which case \(\smash{k_\perp = j_{0,1}\smalldiv a}\) hereafter. (\(\smash{j_{0,1}\approx2.4048}\).) 

Normalizable WPs of the form
\begin{equation}\label{wp}
    \ket{\sm s} = \int_0^{\infty}\!\!\!dk~A(k)\ket{\sm s,k}\kern-0.1em,\qquad s=\uparrow,\sm\downarrow,
\end{equation}
may be built out of the (basis) states \(\smash{\ket{s,k}}\) for suitable choices of \(A(k)\). Since \(\smash{\ket{\sm s,-\sm k}=-\ket{\sm s,k}}\), it suffices to allow only \emph{positive} values of \(k\) in \eqref{wp}. Furthermore, given that the BC \eqref{BBC} is linear in \(\psi\), \(\ket{s}\) satisfies the same BC (in turn, BC \eqref{BC}), and is therefore contained within the WG at all times, as desired. Introducing the integrals
\begin{align}
    F(z,\tau) &= \int_0^{\infty}\!\!\!dk~A(k)\sm\sin(kz)\sm e^{-\sm i\tau\sqrt{k^2+\mu^2}}, \label{F}\\
    G(z,\tau) &= i\!\int_0^{\infty}\!\!\!dk~\frac{A(k)\sm\sin(kz)}{\sqrt{k^2+\mu^2}+1\smalldiv\lambdabar}\sm e^{-\sm i\tau\sqrt{k^2+\mu^2}}, \label{G}
\end{align}
where \(\tau\) has the dimension of length and the constant
\begin{equation}\label{Gam}
   \mu = \sqrt{k_\perp^2+1/\lambdabar^2}
\end{equation}
has that of reciprocal length, Eqs.\ \eqref{wp} can be written as
\begin{subequations}\label{spinupdown}
\begin{align}
 \ket{\uparrow}&=\mqty(J_0(k_\perp\sm\rho)\sm F(z,ct)\sm\mqty(1\\[2pt]0)\\[10pt]-\,J_0(k_\perp\sm\rho)\sm G^\prime(z,ct)\\[5pt] \sm k_\perp J_1(k_\perp\sm\rho)\sm G(z,ct)\sm e^{i\phi}),\\[5pt]   
 \ket{\downarrow}&=\mqty(J_0(k_\perp\sm\rho)\sm F(z,ct)\sm\mqty(0\\[2pt]1)\\[10pt]k_\perp J_1(k_\perp\sm\rho)\sm G(z,ct)\sm e^{-\sm i\phi}\\[5pt] J_0(k_\perp\sm\rho)\sm G^\prime(z,ct)),
\end{align}
\end{subequations}
where \(\smash{G^\prime=\partial_z\sm G}\). Thanks to the unitarity of time evolution, normalizing these WPs within the WG volume at \(\smash{t=0}\), requiring 
\begin{align}\label{normal}
    &\int_0^{\infty}\!\!\!dz~|F(z,0)|^2+k_\perp^2\sm|G(z,0)|^2+|G'(z,0)|^2\nonumber\\
    &\kern4cm =\frac{1}{\pi\sm a^2 J_1^2(k_\perp a)},
\end{align}
ensures their normalization for all \(t\). Here, we used the radial integrals:
\begin{equation*}
    \int_0^a\!\!\!d\rho~\rho\,J_{\nu}^2(k_\perp\sm\rho)\overset{\eqref{rbc}}{=}\frac{a^2}{2}\sm J_1^2(k_\perp a),\quad \nu=0,1.
\end{equation*}
It is convenient to rewrite Eq.\ \eqref{normal} directly in terms of \(A(k)\) as (see \cite{normalize})
\begin{equation}\label{normal2}
    \int_0^{\infty}\!\!\!dk~\frac{\sqrt{k^2+\mu^2}}{\sqrt{k^2+\mu^2}+1\smalldiv\lambdabar}\,|A(k)|^2 =\frac{1}{(\pi a)^2 J_1^2(k_\perp a)}.
\end{equation}
For the left-hand side to be well-defined, \(|A(k)|\) must decay faster (slower) than at least \(1/\!\sqrt{k}\) as \(\smash{k\to\infty\,(0)}\).

For the discussion of BF, we consider a generic, normalized WP \(\ket{\sm \vartheta,\varphi}=\cos(\vartheta\!\smalldiv2)\ket{\uparrow}+e^{i\varphi}\sin(\vartheta\!\smalldiv2)\ket{\downarrow}\) parameterized by the (Bloch sphere) angles \(\smash{\vartheta\in [0,\pi]}\) and \(\smash{\varphi\in[0,2\pi)}\), which includes \(\ket{\uparrow}\) and \(\ket{\downarrow}\) as special cases. Explicitly,
\begin{equation}\label{explicit}
    \ket{\sm \vartheta,\varphi}=\mqty(J_0(k_\perp\sm\rho)\sm F(z,ct)\chi(\vartheta,\varphi)\\[2pt]-\left(\sigma_z\sm\frac{\partial}{\partial z}+\sigma_\rho\sm\frac{\partial}{\partial \rho}\right)\kern-0.1em J_0(k_\perp\sm\rho)\sm G(z,ct)\chi(\vartheta,\varphi)),
\end{equation}
where
\begin{equation}
  \chi(\vartheta,\varphi)=\mqty(\cos\frac{\vartheta}{2}\\[3pt]e^{i\varphi}\sin\frac{\vartheta}{2}),\quad \sigma_\rho=\mqty(0 & e^{-i\phi} \\ e^{i\phi} & 0), 
\end{equation}
and \(\smash{\sigma_z=\text{diag}(1,-1)}\). The large component of \(\ket{\sm \vartheta,\varphi}\) surviving the nonrelativistic limit is said to be ``spin-polarized'' along the unit vector \(\smash{\sin\vartheta\sm(\cos\varphi\Ex+\sin\varphi\Ey)+\cos\vartheta\Ez}\) inclined at an angle \(\vartheta\) to the WG-axis. Given that the WG-axis picks out a preferred direction in this problem, the cases \(\smash{\vartheta=0}\) (\(\pi\smalldiv2\)) denoting spin-polarization parallel (perpendicular) to that axis, display markedly different BF properties.

\paragraph*{A primer on backflow:} Quantum BF naturally arises in the study of scattering experiments, especially arrival-time (or time-of-flight) experiments, in discussions of which BF was recognized early on \cite{Allcock1,*Allcock2,*Allcock3}. In a quantum scattering problem, \(\smash{\vb{j}\cdot d\vb{s}\,dt}\), where \(\vb{j}\) is the probability current density, is suggested to be the differential probability for a particle to be detected on the area element \(d\vb{s}\) of a (oriented) surface \(\mathcal{D}\) between times \(t\) and \(t+dt\) \cite{DDGZ,*DDGZ96}. (This assumption is often implicit in the derivation of the scattering cross-section formula, see \cite[p.\ 912]{CohenT}.) For this probability assignment to be consistent, it must be true that
\begin{equation}\label{CPC}
    \forall\sm t>0,\quad \forall\sm\vb{r}\in \mathcal{D},\quad\vb{j}(\vb{r},t)\cdot d\vb{s} \ge 0.
\end{equation}
This ``current positivity condition'' is seldom explicitly stated, because by the very nature of the free Dirac (or Schr\"odinger) evolution, \eqref{CPC} holds for just about any surface \(\mathcal{D}\) placed in the far-field (i.e., at distances large compared to the support of the initial WP, as is typical of scattering experiments). This asymptotic current positivity is central to the so-called flux-across-surfaces theorems \cite{Peter,DDGZ,*DDGZ96}. A WP violating \eqref{CPC} is said to display quantum BF \cite{explain}, for which the above differential crossing probability becomes negative, and considerations going well beyond the typical textbook scattering analysis become inevitable. It is for this reason that quantum BF is of much interest.

Given a surface \(\mathcal{D}\) and some reasonable initial wave function, \eqref{CPC} tends to hold even in the presence of certain external potentials \cite[p.\ 123]{DDST}. This explains why BF, although permitted in principle, is intrinsically volatile and/or difficult to induce, except perhaps in the near-field regime. The situation conceivably improves upon letting the WP evolve in the presence of a \emph{long-range} potential (e.g., a WG), as opposed to propagating freely, since in this case the asymptotic form of the WP (more precisely, of the probability current) is considerably modified from its free-evolution counterpart (primarily) responsible for \eqref{CPC}. 

While the WG does play a crucial role in provoking and controlling the amount of BF in our set-up (see below), the distinctive spin-dependent flow characteristic of the Dirac current, Eq.\ \eqref{Dcurrent}, is indispensable for achieving the same. In fact, the BF seen here would manifest neither for a spin-0 particle moving within the WG, nor for a spin-1/2 particle moving freely.

\paragraph*{Persistent spin-dependent backflow:} That being said, we proceed to show that \eqref{CPC} is violated for the WP \(\ket{\sm\vartheta,\varphi}\), Eq.\ \eqref{explicit}, with the cross-sectional plane \(\smash{z=L}\) of Fig.\ \ref{WWGG} serving as \(\mathcal{D}\). First, the component of the Dirac current perpendicular to \(\mathcal{D}\), \emph{viz.}, \(\smash{j_z=c\sm\psi^\dagger\alpha_z\psi}\), for this WP reads
\begin{align}
    j_z&=-\sm 2\, c\sm J_0^2(k_\perp\rho)\sm\text{Re}[F^* G^\prime](z,ct)\sm-\sm2\,c\, k_\perp\sin(\varphi-\phi)\nonumber\\[2pt]
    &\kern1cm\times \sin\vartheta\sm J_0(k_\perp\rho)\sm J_1(k_\perp\rho)\sm\text{Im}[F^*G](z,ct).
\end{align}
(It is instructive to note that the first and second terms correspond in the nonrelativistic limit to the \(z\)-components of the convective and spin current densities of Pauli's equation \cite{Mike}, given by \((\hbar\smalldiv m)\sm\text{Im}[\psi^{\text{NR}\sm\dagger}\bm{\nabla}\psi^{\text{NR}}]\) and \((\hbar\smalldiv 2m)\sm\bm{\nabla}\times(\psi^{\text{NR}\sm\dagger}\bm{\sigma}\psi^{\text{NR}})\), respectively, where \(\psi^{\text{NR}}\) is the two-component nonrelativistic wave function.) Given the smallness of the Compton wavelength \(\smash{\lambdabar \approx3.86\times10^{-13}\sm\text{m}}\), it is optimal to consider \(\smash{L\gg\lambdabar}\), for which this component is given approximately by
\begin{align}\label{asymp}
    j_z(\rho,\phi,L,t) &\sim c\pi\sm\frac{(ct)^2k_{\parallel}^3}{\mu^2L^3}\sm P_A(\rho,L,ct)\sm\big[k_\parallel J_0(k_\perp\rho)\nonumber\\
    &\kern1cm -k_\perp\sin\vartheta\sin(\varphi-\phi)J_1(k_\perp\rho)\big],
\end{align}
where \(\smash{k_\parallel=\mu L\smalldiv\!\sqrt{(ct)^2-L^2}}\), and
\begin{equation}
    P_A(\rho,L,ct) = \frac{J_0(k_\perp\rho)\sm|A(k_\parallel)|^2}{\sqrt{k_{\parallel}^2+\mu^2}+1/\lambdabar}
\end{equation}
is a \emph{positive} prefactor irrelevant for the discussion of BF (which contains all the details of the initial condition via \(A\)). To arrive at this expression, we applied the stationary phase argument \cite[Sec.\ 2.9]{stationaryphase} to \eqref{F} and \eqref{G} treating \(L\smalldiv\lambdabar\) as the large (dimensionless) parameter, \(\smash{ct>L}\), and \(\smash{\lambdabar\mu\!\overset{\eqref{Gam}}{\sim}\!1}\).

\begin{figure}[!ht]
    \centering
    \includegraphics[width=0.7\columnwidth]{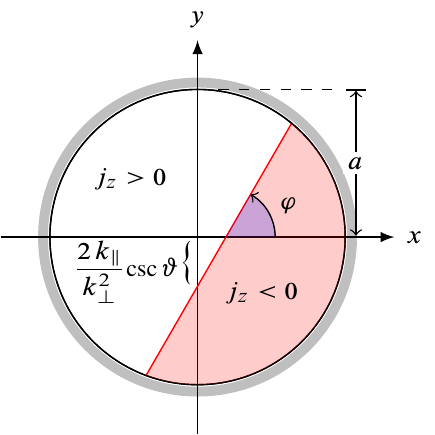}
    \caption{Regions of persistent backflow marked on the cross-section of the waveguide.}
    \label{BFregions}
\end{figure}

The sign of \eqref{asymp} is therefore determined by the quantity enclosed in square brackets, irrespective of \(A(k)\). Noting that both \(\smash{J_{0,1}(k_\perp\rho)\ge0}\) for \(\smash{0\le\rho\le a}\), and \(\smash{\sin\vartheta\ge0}\) since \(\smash{\vartheta\in[0,\pi]}\), it follows that \(\smash{j_z(\rho,\phi,L,t) < 0}\) whenever \(\smash{J_0(k_\perp\rho)\smalldiv J_1(k_\perp\rho)<(k_\perp\smalldiv k_\parallel)\sin\vartheta\sin(\varphi-\phi)}\). However, given that \(\smash{J_0(k_\perp\rho)\smalldiv J_1(k_\perp\rho) < 2\smalldiv (k_\perp\rho)}\) within the WG \cite{Bessel}, a sufficient condition for observing BF is
\begin{equation}\label{chiefineq}
    \rho\sm\sin(\varphi-\phi)>2\sm \csc\vartheta\sm\big(k_\parallel/k_\perp^2\big)=\frac{2\sm a^2\sm(\mu L)\sm\csc\vartheta}{j_{0,1}^2\sqrt{(ct)^2-L^2}}.
\end{equation}
This inequality is graphed in Fig.\ \ref{BFregions}.

Clearly, for \(\smash{\vartheta=0}\) (or \(\pi\)) \eqref{chiefineq} cannot be satisfied owing to \(\csc\vartheta\) becoming unbounded, consequently BF would \emph{not} be observed for WPs spin-polarized parallel (resp.\ anti-parallel) to the WG-axis. For any other \(\vartheta\), especially \(\smash{\vartheta\approx\pi\smalldiv2}\), BF is observed. Once induced, BF remains stable in time since the right-hand side of \eqref{chiefineq} decreases with increasing \(t\). In fact, for \(\smash{t\gg \mu\sm L\smalldiv c}\), \(\smash{k_\parallel\approx 0}\), and \(\smash{j_z<0}\) on roughly half of the cross-section of the WG. Besides, for \(t\) fixed and \(\smash{\vartheta\ne0,\sm\pi}\), BF is controlled by the WG-radius \(a\), given the right-hand side of \(\smash{\eqref{chiefineq}\propto a^2}\). While small \(a\)s favour BF, BF disappears altogether in the limit \(\smash{a\to\infty}\) representing a freely moving particle, in accordance with scattering expectations. A confining WG is therefore essential to observing this intriguing spin-polarization dependent BF.

\paragraph*{Some exact wave packets:} Ideally, we seek solutions with specified initial conditions, e.g., \(F(z,0)\) and \(G(z,0)\) for WPs of the type \eqref{explicit}. However, \(F\) and \(G\) cannot be prescribed independently because they share the same \(A(k)\), cf.\ Eqs.\ (\ref{F}-\ref{G}). Finding compatible \(F\) and \(G\) pairs is a formidable exercise in itself \cite{FGconnection}. Secondly, having arrived at a compatible pair, it is unlikely that their time evolution could be worked out analytically.

Instead, we will find special \(A(k)\)s for which both \eqref{F} and \eqref{G} can be evaluated in closed-form \cite{guess}. We show that \(A(k)\)s of the form
\begin{align}\label{nice}
    A(k) = \frac{A_0}{\sqrt{k^2+\mu^2}}\big(k_{\phantom{\perp}}^2\kern-0.1em+k_\perp^2\big) f(k)
\end{align}
allow us to successfully evaluate Eqs.\ (\ref{F}-\ref{G}) for certain choices of \(f(k)\). For now, \(f(k)\) is arbitrary, limited only by the requirements outlined below Eq.\ \eqref{normal2}, and \(A_0\) is a normalization constant fixed by \eqref{normal2}. (If desired, the \(\sqrt{k^2+\mu^2}\) denominator could be easily jettisoned, as shown below \eqref{KK}.) To see this, employ the above ansatz \eqref{nice} in equation \eqref{F}, arriving at
\begin{align}\label{FF}
    F(z,\tau)=A_0\left(k_\perp^2-\frac{\partial^2}{\partial z^2}\right)\kern-0.1em I(z,\tau),
\end{align}
where
\begin{equation}\label{defI}
    I(z,\tau)=\int_0^{\infty}\!\!\!dk~\frac{f(k)\sin(kz)}{\sqrt{k^2+\mu^2}}\sm e^{-\sm i\tau\sqrt{k^2+\mu^2}}.
\end{equation}
Evaluating \eqref{defI} as it stands is no simpler than evaluating \eqref{F} itself. But a way through comes from the following key step: using \eqref{nice} in \eqref{G} yields, in view of the identity
\begin{equation}
    k_{\phantom{\perp}}^2\kern-0.1em+k_\perp^2 \overset{\eqref{Gam}}{=} \left(\!\sqrt{k^2+\mu^2}+1/\lambdabar\right)\!\left(\!\sqrt{k^2+\mu^2}-1/\lambdabar\right)\!,
\end{equation}
the result
\begin{align}\label{GG}
    G(z,\tau) &= iA_0\!\int_0^{\infty}\!\!\!dk~\frac{f(k)}{\sqrt{k^2+\mu^2}}\left(\!\sqrt{k^2+\mu^2}-1\smalldiv\lambdabar\right)\nonumber\\[-3pt]
    &\kern3.5cm\times\,\sin(kz)\,e^{-\sm i\tau\sqrt{k^2+\mu^2}}\nonumber\\
    &\overset{\eqref{defI}}{=}A_0\left(\frac{1}{i\lambdabar}-\frac{\partial}{\partial \tau}\right)\kern-0.1em I(z,\tau).
\end{align}
Thus, both \(F\) and \(G\) can now be obtained by {\emph {differentiating}} a single function \(I(z,\tau)\).

A few choices of \(f(k)\) lead to exact expressions for \(I\), in turn \(F\) and \(G\). For example, 
\begin{equation}\label{f1}
    f(k) = k\sm e^{-\sm \tau_{\scaleto{0}{3pt}}\sqrt{k^2+\mu^2}}
\end{equation}
with \(\smash{\tau_0>0}\) a free parameter determining the width of the WP at time zero, gives \cite[Eq.\ 3.914.9]{GH}
\begin{equation}\label{I1}
    I(z,\tau) =  \mu z\sm\frac{K_1\!\left(\mu\sqrt{z^2+(\tau_0+i\kern-0.1em\tau)^2}\sm\right)}{\sqrt{z^2+(\tau_0+i\kern-0.1em\tau)^2}};
\end{equation}
in terms of the modified Bessel function of the second kind \(K_\nu\). Relativistic WPs resembling \eqref{I1} have appeared elsewhere \cite{Salpeter,operational,hopfian}, which is not surprising as \(I(z,\tau)\) solves the one-dimensional Klein-Gordon equation with mass \(\mu\). (Of course, not {\emph {every}} solution of the Klein-Gordon equation is of the form \eqref{defI}, see e.g., \cite[Eq.\ (3.20)]{splashpulse}.) Next, letting
\begin{equation}\label{f2}
    f(k) = 2\sin(kz_0)\sm e^{-\sm \tau_{\scaleto{0}{3pt}}\sqrt{k^2+\mu^2}},
\end{equation}
we obtain 
\begin{align}\label{I2}
    I(z,\tau)&= K_0\Big(\mu\sqrt{(z_0-z)^2+(\tau_0+i\kern-0.1em\tau)^2}\sm\Big)\nonumber\\
    &\kern1cm-\sm K_0\Big(\mu\sqrt{(z_0+z)^2+(\tau_0+i\kern-0.1em\tau)^2}\sm\Big),
\end{align}
applying \cite[Eq.\ 3.914.4]{GH}. On the other hand, for
\begin{equation}\label{f3}
    f(k) = \sqrt{\frac{8}{k}}\sin(kz_0)\sm e^{-\sm \tau_{\scaleto{0}{3pt}}\sqrt{k^2+\mu^2}},
\end{equation}
we find
\begin{align}\label{I3}
   I(z,\tau)&=\sqrt{\pi\sm(z_0-z)}\, I_{-1/4}\big(\xi_-(-\,z)\big)\sm K_{1/4}\big(\xi_+(-\,z)\big)\nonumber\\
    &\quad -\sqrt{\pi\sm(z_0+z)} \,I_{-1/4}\big(\xi_-(z)\big)\sm K_{1/4}\big(\xi_+(z)\big),
\end{align}
where \(\xi_\pm(z) = (\mu\kern-0.1em\smalldiv2)\sm\big[\kern-0.1em\sqrt{(\tau_0+i\kern-0.1em\tau)^2\kern-0.1em+\kern-0.1em(z_0+z)^2}\pm(\tau_0+i\kern-0.1em\tau)\big]\), and \(I_\nu\) is the modified Bessel function of the first kind. Eq.\ \eqref{I3} follows from the Laplace transform identity \cite[p.\ 88, Eq.\ (10)]{Prudnikov} after substituting \(\smash{x=\sqrt{k^2+\mu^2}}\) in \eqref{defI}. If \(\sin(kz_0)\) is replaced by \(\sinh(k z_0)\) in Eqs.\ (\ref{f2}-\ref{f3}) with \(\smash{z_0\!<\!\tau_0}\), then letting \(\smash{z_0\mapsto i z_0}\) in Eqs.\ (\ref{I2}-\ref{I3}) and multiplying the results by \(-\,i\) yields the corresponding \(I\). For that matter, \(\sin(kz_0)\) substituted for any antisymmetric periodic function in these examples can also be handled via the Fourier sine-series. As a final example, we take
\begin{equation}
    f(k) = \sqrt{4\sm\pi}\sqrt{k}\sm e^{-\,k z_{\scaleto{0}{3pt}}}.
\end{equation}
In this case,
\begin{align}\label{KK}
    I(z,\tau)&= -\,\frac{\partial}{\partial z}\Big[\sqrt{z_0+iz}\,K_{1/4}\big(i\xi_+\big)\sm K_{1/4}\big(i\xi_-\big)\nonumber\\
    &\kern1cm + \sqrt{z_0-iz}\,K_{1/4}\big(i\xi_+^*\big)\sm K_{1/4}\big(i\xi_-^*\big)\Big],
\end{align}
where \(\xi_\pm\) are the roots of the quadratic equation \(\xi^2-\mu\tau \xi-\mu^2(z_0+iz)^2/4=0\), see \cite[p.\ 42, Eq.\ (1)]{Prudnikov}.

Additional solutions may be obtained by differentiating the above w.r.t.\ different parameters, e.g.\ \(z_0\), and even superposing the results. For instance, differentiating any \(I\) w.r.t.\ \(\tau\) and multiplying the result by \(i\) yields the \(I\) for the same underlying \(A(k)\) without the \(\sqrt{k^2+\mu^2}\) denominator of Eq.\ \eqref{nice}. In the end, the complete WP is obtained via Eqs.\ (\ref{FF}-\ref{GG}), requiring only differentiation of modified Bessel functions easily handled by standard math software like Mathematica or Python. 

It is a pleasure to thank J.\ M.\ Wilkes for carefully reading the manuscript and for helping improve its presentation. Discussions with Markus N\"oth and D.-A.\ Deckert were very valuable.
\bibliography{Ref}

\begin{thebibliography}{97}
\bibitem [{sch()}]{schsolvable}%
  \BibitemOpen
  \href@noop {} {}\bibinfo {note} {Contrast this with the many axisymmetric
  potentials for which Schr\"odinger's equation is solvable:
  \(-1\smalldiv\rho\), \(\rho^2\), renormalized \(\delta(\rho)\),
  \(\rho^2+a\smalldiv\rho^2\), \(\rho+a\smalldiv\rho^2\),
  \(\rho^2+a\sm\rho^{\pm4}+b\sm\rho^{\pm6}\), etc.}\BibitemShut {Stop}%
\bibitem [{not()}]{notation}%
  \BibitemOpen
  \href@noop {} {}\bibinfo {note} {That is, we are equating the abstract state
  vector \(\ket{\,\dotsi\sm}\) to it's representation in the position basis for
  brevity.}\BibitemShut {Stop}%
\bibitem [{par()}]{parity}%
  \BibitemOpen
  \href@noop {} {}\bibinfo {note} {The parity transformation is implemented, up
  to an undetermined phase, by flipping the spatial coordinate and multiplying
  the spinor by \(\beta\). Note that \(\smash{\ket{s,k}\mapsto-\ket{s,k}}\),
  \(\smash{s=\uparrow,\downarrow}\), upon spatial reflection, i.e.,
  \(\smash{(\rho,\phi,z)\mapsto(\rho,\phi+\pi,-z)}\), and simultaneous
  multiplication by \(\beta\) \((\smash{=\text{diag}(1,1,-1,-1)}\) in the
  standard representation).}\BibitemShut {Stop}%
\bibitem [{gue()}]{guess}%
  \BibitemOpen
  \href@noop {} {}\bibinfo {note} {Given the freedom to choose \(A\), one is
  inclined to consider \(A(k)\propto \sqrt{k^2+\mu^2}+1\smalldiv\lambdabar\),
  eliminating the awkward denominator in \eqref{G} and easing the computations
  to some extent. But the resulting WPs may not have a meaningful
  nonrelativistic limit, since \(\lambdabar\) \(\smash{(\mu)\to0}\)
  \((\infty)\) in this limit. Although this is not an issue with regard to
  finding exact solutions, we chose to avoid such solutions.}\BibitemShut
  {Stop}%
\end{thebibliography}
\end{document}